\documentclass[prd,aps,showpacs,amsmath,amssymb,twocolumn,nofootinbib]{revtex4}
\usepackage{graphicx}% Include figure files
\usepackage{dcolumn}% Align table columns on decimal point
\usepackage{bm}% bold math

\begin{document}

%%%%
%    Greek Letters
%

\let\a=\alpha      \let\b=\beta       \let\c=\chi        \let\d=\delta
\let\e=\varepsilon \let\f=\varphi     \let\g=\gamma      \let\h=\eta
\let\k=\kappa      \let\l=\lambda     \let\m=\mu
\let\o=\omega      \let\r=\varrho     \let\s=\sigma
\let\t=\tau        \let\th=\vartheta  \let\y=\upsilon    \let\x=\xi
\let\z=\zeta       \let\io=\iota      \let\vp=\varpi     \let\ro=\rho
\let\ph=\phi       \let\ep=\epsilon   \let\te=\theta
\let\n=\nu
\let\D=\Delta   \let\F=\Phi    \let\G=\Gamma  \let\L=\Lambda
\let\O=\Omega   \let\P=\Pi     \let\Ps=\Psi   \let\Si=\Sigma
\let\Th=\Theta  \let\X=\Xi     \let\Y=\Upsilon

%
%%%

%%%
%    Calligraphic letters
%

\def\cA{{\cal A}}                \def\cB{{\cal B}}
\def\cC{{\cal C}}                \def\cD{{\cal D}}
\def\cE{{\cal E}}                \def\cF{{\cal F}}
\def\cG{{\cal G}}                \def\cH{{\cal H}}
\def\cI{{\cal I}}                \def\cJ{{\cal J}}
\def\cK{{\cal K}}                \def\cL{{\cal L}}
\def\cM{{\cal M}}                \def\cN{{\cal N}}
\def\cO{{\cal O}}                \def\cP{{\cal P}}
\def\cQ{{\cal Q}}                \def\cR{{\cal R}}
\def\cS{{\cal S}}                \def\cT{{\cal T}}
\def\cU{{\cal U}}                \def\cV{{\cal V}}
\def\cW{{\cal W}}                \def\cX{{\cal X}}
\def\cY{{\cal Y}}                \def\cZ{{\cal Z}}

\def\dbd{{$0\nu 2\beta\,$}}
%
%%%%

\newcommand{\Ns}{N\hspace{-4.7mm}\not\hspace{2.7mm}}
\newcommand{\qs}{q\hspace{-3.7mm}\not\hspace{3.4mm}}
\newcommand{\ps}{p\hspace{-3.3mm}\not\hspace{1.2mm}}
\newcommand{\ks}{k\hspace{-3.3mm}\not\hspace{1.2mm}}
\newcommand{\des}{\partial\hspace{-4.mm}\not\hspace{2.5mm}}
\newcommand{\desco}{D\hspace{-4mm}\not\hspace{2mm}}

%%%%

%\draft command makes pacs numbers print
%\draft

\title{\boldmath BICEP2, non Bunch-Davies and Entanglement}

\author{Namit Mahajan
}
\email{nmahajan@prl.res.in}
\affiliation{
 Theoretical Physics Division, Physical Research Laboratory, Navrangpura, Ahmedabad
380 009, India
}

%\date{\today}

\begin{abstract}
BICEP2 result on the tensor to scalar ratio $r$ indicates a blue tilt in the primordial gravitational wave
spectrum. This blue tilt and the observed large value $r = 0.2$ are difficult to accommodate within the single field
inflationary scenarios under standard conditions. Non Bunch-Davies vacuum states have been proposed as a possibility.
Such vacuua are known to lead to pathologies. In this note we propose to interpret them as excited/squeezed states built 
over the standard Bunch-Davies vacuum and discuss the associated entanglement properties due to de Sitter horizon, and how
such an approach may be more natural in the context of inflation.
\end{abstract}

\pacs{
}
\maketitle
%\narrowtext

%\section{Introduction}

In the last one decade or so cosmology has entered the precision era with WMAP \cite{Hinshaw:2012aka}
and PLANCK \cite{Ade:2013uln} dramatically improving upon
the COBE observations \cite{Mather:1993ij} and providing an accurate set of measurements for other cosmological parameters.
Very recently, BICEP2 announced the result on the measurement of the ratio of tensor to scalar spectrum, $r$ \cite{Ade:2014xna}.
The quoted result for $r$ is $0.2^{+0.07}_{-0.05}$ (with foreground subtraction the central value reads $r=0.16$). 
If the BICEP result
is confirmed by other experiments, this
would constitute a direct signature of primordial gravitational waves. PLANCK collaboration had provided
with an upper limit on $r$ \cite{Ade:2013uln}, 
though at different $\ell$ value than that of BICEP2. There is therefore some tension between the two values.

Inflationary paradigm \cite{Guth:1980zm} predicts the power spectra
of both the scalar and the gravitational amplitudes (see for example \cite{Baumann:2009ds} for a quick introduction to the basics
and relevant expressions). However, a large value of $r$ as measured by the BICEP2 collaboration
is difficult to achieve in the usual single filed slow roll models. Typically, these models predict that the 
scalar spectral index $n_s-1 = d\ln P_S/d\ln k \sim -2\epsilon$ while the tensor spectral index 
$n_T = d\ln P_T/d\ln k = -2\epsilon$.
There is an additional 'consistency relation', $r= 16 \epsilon = -8 n_T$, which ties the two in a simple but strong fashion.
The main difficulty, apart from satisfying the large value of $r$, is the fact that 
an inflationary model should predict a positive (implying a blue tilt for the gravitational wave
spectrum as opposed to the observed red tilt for the scalar power spectrum) and large $n_T$ which is is 
hard to achieve in simple models. 
Attempts have been made to reconcile the BICEP2 result and the PLANCK upper limit by relaxing some of the 
conditions/constraints while fitting to the data sets \cite{Cheng:2014bma}.
%Theories which violate null energy condition (for a quick introduction see \cite{Rubakov:2014jja})
%can lead to a 
%blue tilt for the gravitational waves. String gas cosmology is one example which is shown to be consistent
%with the BICEP2 results \cite{Brandenberger:2014faa}. 
Another interesting possibility is to consider departure from the usual 
Bunch-Davies vacuum to get enhanced spectra \cite{Collins:2014yua} (see also \cite{Hui:2001ce} for some early work). This will the focus of this note.

We begin by first recalling some of the
properties and pathological implications of non Bunch-Davies vacuum states (for the issues about
quantum field theory in de Sitter space that are
relevant for the present discussion see \cite{Bousso:2001mw}). For simplicity consider
a scalar filed $\varphi$ in the de Sitter space. The arguments can be straightforwardly carried over to the graviton.
It is known that there is no unique vacuum state in a general curved space-time. Given a quantum field, one expands it in terms of a complete set as
\begin{equation}
 \varphi = \sum_n [a_n \varphi_n^{(+)} + a_n^{\dag}\varphi_n^{(-)}] \label{bd}
\end{equation}
where $a_n$ and $\varphi_n^{(+)}$ are the annihilation operators and positive energy mode functions. Since there is
no unique vacuum state, one could expand the same field in terms of a different set of mode functions:
\begin{equation}
 \varphi = \sum_n [\tilde{a}_n \tilde{\varphi}_n^{(+)} + \tilde{a}_n^{\dag}\tilde{\varphi}_n^{(-)}] \label{alpha}
\end{equation}
The two sets are related to each other via the Bogoliubov transformations:
\begin{equation}
 \tilde{\varphi}_n^{(+)} = \varphi_n^{(+)} \cosh\alpha_n + \varphi_n^{(-)}e^{i\beta_n}\sinh\alpha_n \label{bog}
\end{equation}
or in terms of the creation and annihilation operators\footnote{A general Bogoliubov transformation takes the form:\\
$\tilde{a}_n = A^*_{nn'}a_{n' }- B^*_{nn'}a_{n'}$ with the condition $\vert A\vert^2-\vert B\vert^2=1$.}
\begin{equation}
\tilde{a}_n = a_n \cosh\alpha_n - a_n^{\dag}e^{-i\beta_n}\sinh\alpha_n
\end{equation}
This again displays the fact that the positive frequency modes do not remain positive frequency modes unless
$\alpha_n$ are chosen such that the second term above vanishes. In the de Sitter space-time, the common choice
is the so called Euclidean or the Bunch-Davies vacuum (sometimes also referred to as thermal vacuum) \cite{Bunch:1978yq}. 
In this case,
one solves the equation of motion (in conformal time, $\eta$) and then demanding that the solution has the correct
Minkowski form for high frequencies as $\eta\to -\infty$, one picks out one of the two independent solutions. The end result
is (for $\eta\to -\infty$)
\begin{equation}
 \varphi_k^{(+)} = H^{(2)}(k\eta) \rightarrow \left(\frac{2}{\pi k\eta}\right) e^{-ik\eta}
\end{equation}
and $ \varphi_k^{(-)} = (\varphi_k^{(+)})^* = H^{(1)}(k\eta)$
where $H^{(1,2)}$ are the Hankel functions.
We therefore have two vacuua: $a_n\vert E\rangle  = 0$ and $\tilde{a}_n\vert\alpha\rangle = 0$. 
In the notation we follow, $\vert E\rangle$ denotes the standard Bunch-Davies or Euclidean vacuum 
and $\vert\alpha\rangle$ denotes the other vacuum choice, called $\alpha$-vacuua \cite{Mottola:1984ar}. $\alpha = 0$
leads to the Euclidean vacuum. It is known that
Eq.(\ref{bd}) respects the de Sitter invariance of the background space-time, while for the Eq.(\ref{alpha}) this 
happens only if $\alpha_n = \alpha$ (i.e. n-independent). We therefore choose to work with this choice. Further,
$\beta_n \neq 0$ in Eq.(\ref{bog}) implies time reversal non-invariance (and also CPT). 
We thus set $\beta_n$ to zero. 
The Euclidean vacuum is easily obtained from the Euclidean de Sitter i.e. 
sphere via analytic continuation and results in the well known two point functions
\begin{equation}
 G_E(x,x') = \langle E\vert\varphi(x)\varphi(x')\vert E\rangle
 \propto\,  _2F_1\left(h_+,h_-,\frac{d}{2},\frac{1+Z}{2}\right) 
\end{equation}
where $_2F_1$ is the hypergeometric function, $d$ is the space-time dimensionality, 
$h_{\pm} = (d-1)/2\pm i\nu$ with $\nu = \sqrt{m^2 - \left(\frac{d-1}{2}\right)^2}$ and $Z$ is the invariant
distance between two points. $Z(x,x')$ is greater (less) than unity for time-like (space-like) intervals
while its is unity for light-like separations. In the above equation, we have considered the two point
function (Wightman function) but we shall interchangeably talk about the Hadamard function $G^{(1)}$ which is 
nothing but the vacuum expectation of the anti-commutator. This function captures the state dependence.
One can now evaluate Hadamard function $G^{(1)}_{\alpha}$ in terms of the Euclidean modes
\begin{equation}
 G^{(1)}_{\alpha}(x,x') = \cosh 2\alpha G^{(1)}_E(Z) + \sinh 2\alpha G^{(1)}_E(-Z)
\end{equation}
which shows that the $\alpha$ vacuum function not only depends on $Z$ but also on the antipodal
point $-Z$. This feature shows up in other two point functions like the Feynman propagator as well and it is
this which is at the heart of pathologies. Specifically, the two point functions have singularities at the 
antipodal points. Such points are separated by a horizon \footnote{Recall that the de Sitter space has a horizon
and also carries Gibbons-Hawking temperature $T_{dS} = H/(2\pi)$ \cite{Gibbons:1977mu}, where the Ricci scalar $R=12 H^2$.}. 
Moreover, $\alpha$ vacuum two point functions have non-Minkowski analytic structure 
in the coincident limit, coming via the contributions due to negative frequency modes. These observations
already raise serious doubts about interpreting $\alpha$ vacuua as sensible vacuum states. The difficulty
with this becomes more serious in an interacting field theory.

Before proceeding further, let us introduce an equivalent way of representing the $\alpha$ modes:
\begin{equation}
 \tilde{\varphi}_n^{(+)} = N_{\alpha'}[\varphi_n^{(+)}  + \varphi_n^{(-)} e^{\alpha'}] \label{bog2}
\end{equation}
where $N_{\alpha'} = 1/\sqrt{1-e^{\alpha'+\alpha'^*}}$ and $\alpha'$ a complex parameter with
$Re(\alpha') < 0$ and the Euclidean vacuum is obtained for $Re(\alpha') \to -\infty$. Again, a complex
$\alpha'$ will lead to troubles with CPT and thus we choose it real.
The two representations are related to each other via the variable transformation $e^{\alpha'} = \tanh\alpha$.

As already discussed above, the $\alpha$ vacuum correlation functions have peculiar properties even in the free
field theory limit and one would not want to consider them any further. Before doing so, it is worthwhile
to have some more check. Consider an Unruh detector moving along a time-like geodesic and
with a linear coupling
to the field \cite{Unruh:1976db}. Denoting by ${\mathcal{D}}$ the operator which acts on the states of the detector 
($\vert{\mathcal{E}}_i\rangle$),
the detector-field coupling is of the form $\int dt\, \varphi(t){\mathcal{D}}(t)$. The transition rate between
two states of the detector can be evaluated to be
\begin{eqnarray}
 {\mathcal{R}}_{\alpha}^{ij} &=& {\mathcal{R}}_{\alpha}^{{\mathcal{E}}_i\to{\mathcal{E}}_j} \nonumber \\
 &=& \vert\langle{\mathcal{E}}_j\vert{\mathcal{D}}\vert{\mathcal{E}_i}\rangle\vert^2 \int dt 
 e^{-i({\mathcal{E}}_j-{\mathcal{E}}_i)t}G_{\alpha}(x(t),x(0))
\end{eqnarray}
Using the explicit form (and properties) of $G_{\alpha}$ it is straightforward to convince oneself that
only for the choice of the Euclidean vacuum i.e. $\alpha =0$ (or equivalently $\alpha'=-\infty$), one has
\begin{equation}
\frac{{\mathcal{R}}_{\alpha=0}^{ij}}{{\mathcal{R}}_{\alpha=0}^{ji}} = e^{-2\pi({\mathcal{E}}_j-{\mathcal{E}}_i)/H} 
\end{equation}
which is the expected thermal behaviour and 
gives the associated temperature to be $T = H/(2\pi)$ in conformity with the Gibbons-Hawking temperature
of the de Sitter space. For any other choice of $\alpha$ one therefore has a general result that the detailed
balance may not work. This observation again singles out the Euclidean or the Bunch-Davies vacuum
as a preferred and unique choice.

Let us next ask if there is a way to interpret the $\alpha$ vacuua. The Bogoliubov transformation Eq.(\ref{bog})
can be formally thought of as achieved via the action of a unitary operator (this is all familiar from quantum
optics where squeezed states are constructed by the action of such unitary operators \cite{quantopt}):
\begin{equation}
S(\xi) = \exp[\sum_n\frac{1}{2} (\xi a_n^{\dag 2} - \xi^*a_n^2)]
\end{equation}
such that $S(\xi)a_nS(\xi)^{\dag} = \tilde{a}_n$. In the $\alpha'$ representation, $\xi = \frac{1}{4}\ln[\tanh(-\alpha'/2)]$. Now,
instead of trying to interpret the $\alpha$ vacuua as genuine vacuum states, one could take the view point that
these are excited states built over the Euclidean vacuum.\footnote{Strictly speaking, for $\alpha_n = \alpha$ i.e.
independent of $n$, the overlap between the vacuum and the $\alpha$ state is zero. This orthogonality
would therefore forbid their interpretation as excited states. However, in an effective theory sense, one can
still have this description allowed and this is the standpoint what we adopt here.} Given these
excited/squeezed states, it is then straightforward to evaluate various correlators and one hopes that
various pathological issues disappear. However, once again we are confronted with the situation that the 
expectation value of the two point (and higher point) functions of the quantum field in these states $\vert \alpha\rangle$
depends on the antipodal point leading to non-local behaviour. This is possible if the states of the system
are entangled leading to EPR \cite{Einstein:1935rr} like effects.

We now ask the question if such a thing is plausible or does one have to simply assume it if these
states are to be put to any use. Since the de Sitter space, like a black hole,  has a horizon, the space naturally
gets divided into two regions. This therefore implies that given a quantum field, 
there will be non-trivial correlations
between the modes inside and outside the horizon. The modes outside (inside) the horizon are not accessible to an
observer within (outside) the horizon and therefore the natural approach would be to trace the modes outside (inside) the horizon
and use the reduced density matrix thus obtained. With an appropriate choice of $\alpha$ (or $\alpha'$), the
desired squeezed state $\vert\alpha^{(')}\rangle$ can therefore be constructed. This state is clearly
not a pure state but
a mixed state and therefore we can expect non-local EPR like effects. In the context of 
inflation, one can think of the scale
at which a given mode exits the horizon: $k_* = aH$. The Fourier modes of the quantum
field now naturally get divided in two sets - 
modes with $k<k_*$ and $k>k_*$. In an interacting field theory 
(interactions can be in the matter sector itself like
$\varphi^4$ terms or can eventually arise due to gravity), this will lead to momentum 
space entanglement between these two
set of modes. The end result, qualitatively, is a mixed state after tracing out one set of the modes,
but now in the momentum space.
Consider two modes - one inside and one outside the horizon and attach comoving observers with both. 
The two observers
feel a relative constant acceleration between them. Approximating the near horizon geometry by 
the Rindler geometry, one can
easily obtain the reduced density matrix corresponding to the state in Region I of the Rindler
space-time by tracing over the states
in the causally disconnected Region II. The reduced density matrix reads (see for example \cite{Crispino:2007eb})
\begin{eqnarray}
\rho_I &\sim& \frac{1}{\cosh^2\alpha}\sum_n \tanh^{2n}\alpha\,\,\, \vert n\rangle^I\langle n\vert^I \nonumber \\
&=& (1-e^{-2\pi\omega/H})\sum_n (e^{-2\pi\omega/H})^n\, \vert n\rangle^I\langle n\vert^I
\end{eqnarray}
The above is a thermal state with Unruh temperature $T_{Unruh} = H/(2\pi) = T_{dS}$. 
This is correct up to terms
${\mathcal{O}}(\alpha=\tanh^{-1}[e^{-\pi k/(aH)}])$ (reason for '$\sim$' in the 
first line of the above equation) which are rather small for $k\geq aH$, where $\omega$ has been traded off
in favour of $k$ and now expressed as physical momentum.
The state of the field in the infinite past is assumed to be the vacuum state which  
corresponds to the two mode squeezed 
state (as in quantum optics) in the late future.

A squeezed state can be constructed by acting the coherent state displacement operator 
$D(\alpha)$ followed by the squeezing operator $S(\xi)$
on the vacuum state $\vert E\rangle$. Since for the harmonic potential 
the coherent state stays a coherent state under time evolution (only the complex parameter changes by a phase),
it is not difficult to establish that a squeezed state also remains a squeezed state with the 
only change that the parameters $\alpha$ and $\xi$
pick up phase factors with phases $\omega t$ and $2\omega t$ respectively. This conclusion 
is expected to change when non-trivial interactions
are considered. The detailed investigation of effect of interactions is beyond the scope of 
this note and we leave it to a separate study \cite{namit}. For the time being
we consider a theory with two fields with some reasonable interactions, both self interactions
and with each other.  In this toy theory, the future
squeezed state is now going to be rather complicated and will depend on the creation and 
annihilation operators 
of the two fields. On generic grounds there is no reason for the two point functions of these 
two fields evaluated in this complicated squeezed
state to be identical. It should therefore be not surprising that the power spectra of the 
two fields have different momentum dependence. 
In particular, it may be possible, at least in a toy theory with tuned interactions, that
power spectrum of one field is nearly scale invariant while
the other field has a power spectrum with a positive scale dependence.

All what has been discussed above can now be applied to the case of inflation. 
In the single field inflation, there are two fields: the inflaton field, a scalar with
a potential which in principle can have self-interactions,  and the graviton field,
a symmetric transverse traceless tensor. In the language of the toy theory
considered above, the inflaton field is the one whose power spectrum is nearly
scale invariant ($n_S-1= -2\epsilon$) 
while the graviton has a blue tilted power spectrum with $n_T \sim 1$ and $r$
consistent with the PLANCK limit.\footnote{Whether it is possible 
to have such a sce2nario realised with a realistic inflaton potential remains
to be established. For the present purpose we assume that it does realise.}
This scenario has been shown
to be consistent with the cosmological data, including the BICEP2 results 
when the $in$ vacuum is chosen to be the Euclidean vacuum
and the $out$ vacuum is chosen to be the one corresponding to a static observer
(see S. Mohanty and A. Nautiyal in \cite{Collins:2014yua}). 
This corresponds to the identification $\alpha' = -\pi k/(aH)$.
As we have discussed above, instead of interpreting as a vacuum state, it is more
appropriate to view the transformed state as a squeezed state. Another advantage
in considering these states as excited states rather than alternate vacuum states is that
by definition a vacuum state would have to be annihilated by the annihilation operators of both the 
fields in the toy theory. Therefore one would have to invoke some mechanism by which
the power spectrum of only one of them gets significant corrections. As argued above,
the squeezed state interpretation has the potential to naturally address this aspect. 

One therefore notes that the discussion and the construction outlined above provides
strong theoretical basis to scenarios where correlation
functions are evaluated in more general states. The states, though, can not be chosen
completely arbitrarily and are supposed to be constructed
following a well defined set of rules. General mixed states have been considered on
a somewhat phenomenological level. They should be
subjected to the checks dictated by the prescribed set of rules. It is still possible
to have multiple such states which turn out to be consistent with
the data at present. As an example, consider a different scenario 
scenario consistent with data where the scalar spectrum also has a non-negligible momentum dependence
together with the tensor spectrum. As we have argued above, it may be possible
to achieve such a scenario by appropriately choosing the
potential. The question to be addressed next is which one of the two
describes our universe. The answer and the capability to
differentiate lies in the higher point correlation functions and
therefore future measurements or limits on the non-Gaussianities will be
essential in settling this issue.

In this note we have focused our attention on the $\alpha$ vacuum states
and their interpretation in the light of the recent results
on the primordial gravitational wave observations by BICEP2 collaborations.
Non Bunch-Davies/Euclidean vacuum states
have been proposed as a solution to the observed blue tilt of the tensor power
spectrum and further to reconcile the positive large
value of $r$ as announced by the BICEP2 and the PLANCK upper limit on $r$.
From the point of view of quantum field theory, 
$\alpha$ vacuua lead to various pathologies like undesirable singularity structure of
the correlation functions and departure from
the standard thermal behaviour of an Unruh detector. Interpreting these as
excited states is an interesting possibility
and we have argued that such an interpretation seems to be quite natural.
It naturally leads to mixed states and this is 
consistent with the presence of the de Sitter horizon. We have further pointed
out that it may be possible to construct potentials and interactions,
albeit fine tuned,  which can lead to desired features in the scalar and tensor
power spectra. Whether in practice it is easy
to realise such a construction remains to be explored.
It is worth mentioning that on the theoretical side, there have been recent 
studies on computing entanglement entropy
in de Sitter space both in the Euclidean \cite{Maldacena:2012xp} and $\alpha$ vacuua \cite{Kanno:2014lma}.
It is found that the entanglement entropy increases monotonically with the value of $\alpha$.
It would be interesting to
consider an interacting theory
and compute the entanglement entropy in the squeezed states. Such a theoretical exercise
may perhaps shed some further
light on the nature of these states in such a context. A particularly relevant quantity would be the
entanglement entropy in the momentum space and 
its scaling properties. This will be investigated elsewhere \cite{namit}.

\end{document}